\documentclass{egpubl}
\usepackage{pg2019}
 
\SpecialIssuePaper         

\usepackage[T1]{fontenc}
\usepackage{dfadobe}  

\usepackage{cite}  
\BibtexOrBiblatex
\electronicVersion
\PrintedOrElectronic
\ifpdf \usepackage[pdftex]{graphicx} \pdfcompresslevel=9
\else \usepackage[dvips]{graphicx} \fi

\usepackage{egweblnk} 

\usepackage{eucal}

\usepackage{breakcites}
\usepackage{ellipsis} 
\usepackage[mathletters]{ucs}
\usepackage[utf8x]{inputenc}
\usepackage{mathtools}
\usepackage{amsbsy}
\usepackage{upgreek}
\usepackage{dblfloatfix}
\usepackage{microtype}

\usepackage[table]{xcolor}

\usepackage{tabularx}
\usepackage{booktabs}

\usepackage{amsmath}

\usepackage{amsfonts}
\usepackage{amssymb}

\definecolor{lightbluishgrey}{rgb}{0.76078,0.88235,0.92157}

\newcommand{\newhl}[1]{{{#1}}}

\usepackage{layouts}

\usepackage{wrapfig}

\newcommand{\refequ}[1] {Equation~(\ref{equ:#1})}

\newcommand{\reffig}[1] {Fig.~\ref{fig:#1}}

\makeatletter
\def\reffig{\@ifnextchar[{\@myreffigloc}{\@myreffignoloc}}
\def\@myreffigloc[#1]#2{Fig.~\ref{fig:#2}, \emph{#1}}
\def\@myreffignoloc#1{Fig.~\ref{fig:#1}}
\makeatother

\newcommand{\refsec}[1] {Section~\ref{sec:#1}}

\let\mat = \mathbf
\usepackage{xfrac}

\usepackage{amsmath}
\usepackage{amssymb}    
\usepackage{cancel}
\usepackage[T1]{fontenc}

\newcommand{\R}{\mathbb{R}}

\newcommand{\vc}[1]{\mathbf{#1}}
\newcommand{\C}{\mat{C}}

\newcommand{\e}{\vc{e}}

\newcommand{\vv}{\vc{p}}

\newcommand{\w}{\vc{w}}
\newcommand{\x}{\vc{x}}

\renewcommand{\C}{\mathcal{C}}

\newcommand{\E}{\mathcal{E}}

\renewcommand{\H}{\mathcal{H}}
\newcommand{\I}{\mat{I}}
\newcommand{\J}{\mathcal{J}}

\newcommand{\RR}{\mat{R}}

\newcommand{\T}{\mat{T}}
\newcommand{\U}{\mathcal{U}}
\newcommand{\V}{\mathcal{V}}

\usepackage{graphicx}

\usepackage{wrapfig}

\makeatletter
\let\save@mathaccent\mathaccent
\newcommand*\if@single[3]{%
  \setbox0\hbox{${\mathaccent"0362{#1}}^H$}%
  \setbox2\hbox{${\mathaccent"0362{\kern0pt#1}}^H$}%
  \ifdim\ht0=\ht2 #3\else #2\fi
  }
\newcommand*\rel@kern[1]{\kern#1\dimexpr\macc@kerna}
\newcommand*\widebar[1]{\@ifnextchar^{{\wide@bar{#1}{0}}}{\wide@bar{#1}{1}}}
\newcommand*\wide@bar[2]{\if@single{#1}{\wide@bar@{#1}{#2}{1}}{\wide@bar@{#1}{#2}{2}}}
\newcommand*\wide@bar@[3]{%
  \begingroup
  \def\mathaccent##1##2{%
    \let\mathaccent\save@mathaccent
    \if#32 \let\macc@nucleus\first@char \fi
    \setbox\z@\hbox{$\macc@style{\macc@nucleus}_{}$}%
    \setbox\tw@\hbox{$\macc@style{\macc@nucleus}{}_{}$}%
    \dimen@\wd\tw@
    \advance\dimen@-\wd\z@
    \divide\dimen@ 3
    \@tempdima\wd\tw@
    \advance\@tempdima-\scriptspace
    \divide\@tempdima 10
    \advance\dimen@-\@tempdima
    \ifdim\dimen@>\z@ \dimen@0pt\fi
    \rel@kern{0.6}\kern-\dimen@
    \if#31
      \overline{\rel@kern{-0.6}\kern\dimen@\macc@nucleus\rel@kern{0.4}\kern\dimen@}%
      \advance\dimen@0.4\dimexpr\macc@kerna
      \let\final@kern#2%
      \ifdim\dimen@<\z@ \let\final@kern1\fi
      \if\final@kern1 \kern-\dimen@\fi
    \else
      \overline{\rel@kern{-0.6}\kern\dimen@#1}%
    \fi
  }%
  \macc@depth\@ne
  \let\math@bgroup\@empty \let\math@egroup\macc@set@skewchar
  \mathsurround\z@ \frozen@everymath{\mathgroup\macc@group\relax}%
  \macc@set@skewchar\relax
  \let\mathaccentV\macc@nested@a
  \if#31
    \macc@nested@a\relax111{#1}%
  \else
    \def\gobble@till@marker##1\endmarker{}%
    \futurelet\first@char\gobble@till@marker#1\endmarker
    \ifcat\noexpand\first@char A\else
      \def\first@char{}%
    \fi
    \macc@nested@a\relax111{\first@char}%
  \fi
  \endgroup
}
\makeatother

\usepackage[ruled,vlined]{algorithm2e}
\usepackage{etoolbox}
\usepackage{array}

\newcommand{\figs}{}
\def\figs/{figs/}

\DeclareUnicodeCharacter{8214}{\|}
\DeclareUnicodeCharacter{188}{{\tfrac{1}{4}}}
\DeclareUnicodeCharacter{189}{{\tfrac{1}{2}}}
\DeclareUnicodeCharacter{8532}{{\tfrac{2}{3}}}
\DeclareUnicodeCharacter{8531}{{\tfrac{1}{3}}}
\usepackage[verbose]{newunicodechar}
\DeclareUnicodeCharacter{0923}{\boldsymbol{\Lambda}}

\newcommand{\ijs}{_{\{i,j\}}}

\begin{document}

\teaser{
  \vspace*{-0.7cm}
  \includegraphics[width=\linewidth]{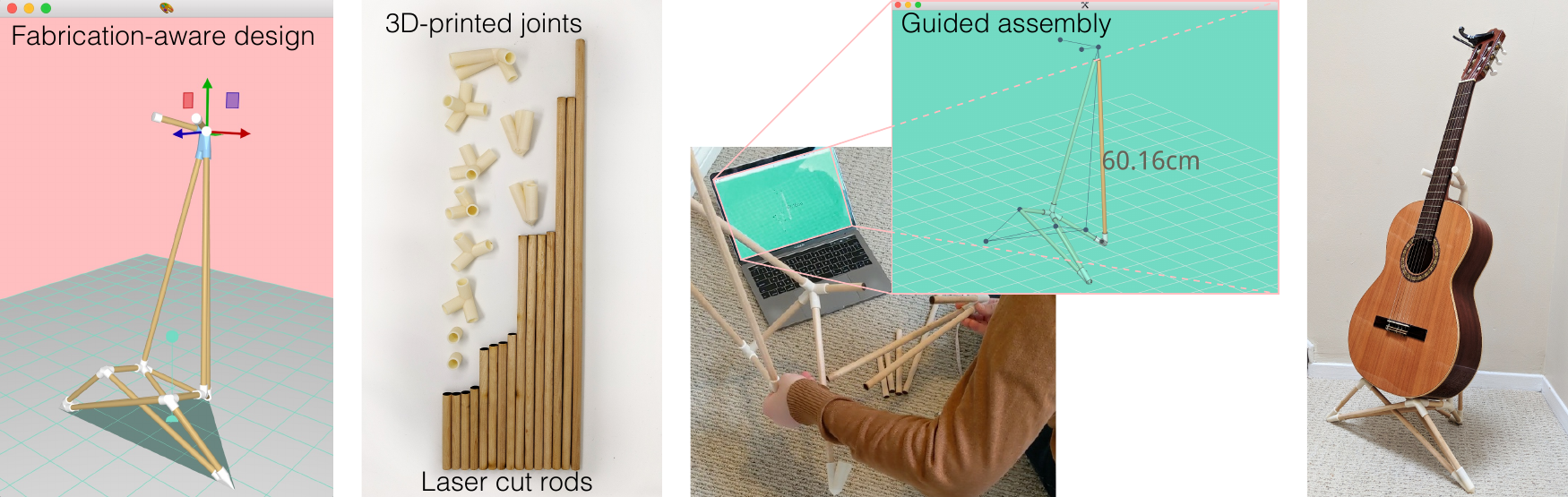}
  \caption{
    \label{fig:teaser}
    \emph{RSDesigner}, our fabrication-aware interface for joint-rod
    structures, helps a user design a \newhl{custom-fit} guitar
    stand with a \newhl{unique aesthetic discovered during modeling}.
    Joint geometries are 3D-printed and wooden dowels are laser cut to length.
    The user physically assembles the structure guided by an on-screen aid,
    \emph{RSAssembler}.
  }
}

\title{
  RodSteward: A Design-to-Assembly System for Fabrication using 3D-Printed
  Joints and Precision-Cut Rods
  \vspace*{-0.7cm}
}

\author{Alec Jacobson, University of Toronto
}

\maketitle

\begin{abstract}
  We present \emph{RodSteward}, a design-to-assembly system for creating
furniture-scale structures composed of 3D-printed joints and precision-cut
rods.
The RodSteward systems consists of: \emph{RSDesigner}, a
fabrication-aware design interface that visualizes accurate geometries during
edits and identifies infeasible designs; physical fabrication of parts via novel
fully automatic construction of solid 3D-printable joint geometries and
automatically generated cutting plans for rods; and \emph{RSAssembler},
a  guided-assembly interface that prompts the user to place parts in order while
showing a focus+context visualization of the assembly in progress.
We demonstrate the effectiveness of our tools with a number of example
constructions of varying complexity, style and parameter choices.

\end{abstract}

%

\section{Introduction}
\label{sec:intro}
Advanced manufacturing processes dramatically increase the complexity of
physically fabricable geometries.
For example, a 3D printer can directly fabricate an intricate, high genus shape,
so long as it fits in the machine's build volume.
In contrast, standard laser cutters have a much larger albeit two-dimensional
cutting bed.
Unfortunately, these complementary strengths are not easily leveraged
harmoniously in a single design.
In particular, large three-dimensional objects are not well suited for either
process in isolation.
Further complicating design, construction of fabricable parts is a non-trivial
task, often requiring slow iterations between virtual design and physical
prototyping.
For example, a design that \emph{looks} feasible, may turn out to have
overlapping or corrupted geometry.

In response, we present the \emph{RodSteward} system, a design-to-assembly
system for creating furniture-scale structures composed of 3D-printed joints and
precision-cut rods (see \reffig{teaser}).
This design space is especially interesting because nearly all geometric
complexity is shifted onto the small joint shapes, harmonizing with the
qualities of the 3D printer.
Meanwhile, the long rods can be purchased \emph{en masse} at any hardware store
and diced up with any tool capable of simple-but-precise perpendicular cuts
(e.g., a laser cutter, but also a handsaw and miter box).
Sparse, wireframe designs are also a currently trendy modern furniture
aesthetic.

The RodSteward system has three stages: 1) \emph{RSDesigner}, a
fabrication-aware design interface; 2) part geometry realization and physical
printing and cutting; and 3) \emph{RSAssembler}, a guided-assembly
interface.
RSDesigner allows the user to edit a virtual structure while
interactively maintaining an accurate visualization of the fabricated parts.
\newhl{Our emphasis on real-time feedback allows a user to fine-tune and evaluate
designs on-the-fly.}
The interface will highlight and alert the user to potential problems with the
design such as overlapping parts or structurally unstable designs.
Complementing this interface, we propose a novel joint geometry construction
algorithm, which generates solid, watertight and 3D-printable joints
given the user's rod-joint network description. 
\newhl{A subset of existing methods} require manual intervention to generate
the literal geometry of each joint, \newhl{breaking the exploratory design
loop.}
\newhl{In contrast, our method belongs to the class of automatic methods with a
tight design loop that} allows the user to focus on the high-level creative
aspects of the overall design, rather than the geometry of each joint itself.
\newhl{
With varying physical accuracy, some previous methods have simulated
the structural stability of rod-joint structures.
Our contributions complement this particular well-explored feature, and we
therefore leave incorporating this aspect as an incremental improvement to
RodSteward.
Instead, we focus on first-order design issues such as rod-intersections and
balance.} 

\begin{figure}
  \includegraphics[width=\linewidth]{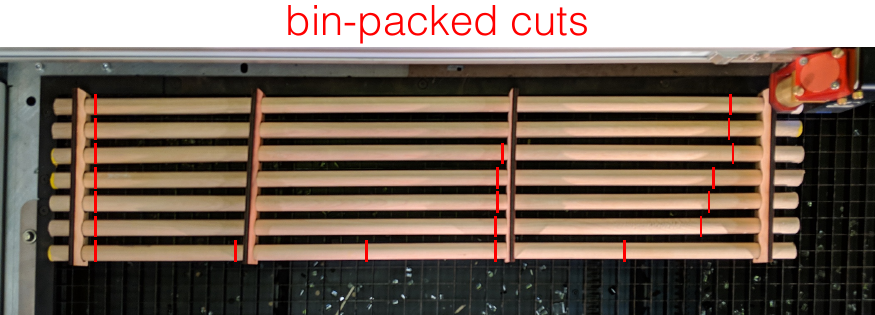}
  \caption{\label{fig:laser-bin-packing} We pack all of the rod-lengths into a
  single cut plan over a small number of raw rods. Rods are positioned in place
  using a ``comb'' jig with holes cut at regular intervals matching the spacing
  of the automatically generated cut plan.}
\end{figure}

Upon design completion, we automatically engrave each joint with a visible I.D.
and send the parts to the 3D-printer.
For rods, we generate a cutting plan that packs the segments into a minimal
number of standard-size rods, so all can be cut in a single, quick job (see
\reffig{laser-bin-packing}).
After fabricating the individual parts, RSAssembler visualizes the
partial structure as the user places each part.
The user presses a hotkey to advance and the guide suggests the next part to
place and updates the visualization.
Without the guided assembly interface, assembling these complex structures would
reduce to solving a frustratingly difficult 3D puzzle.

We demonstrate the effectiveness of RodSteward as a design-to-assembly system,
by constructing 
structures (e.g., \reffig{wolf-head}) that highlight the simplicity and
generality of our system to accommodate
non-manifold edge-networks,
circular and polygonal rod profiles,
acute angles between adjacent rods,
and 
complex yet non-self-intersecting and balancing structures.

%

\section{Related Work}
\label{sec:related-work}
In the past decades, designers, researchers and hobbyists alike have looked for
ways to leverage the geometric complexity afforded by 3D printing with
traditional or unconventional parts.
We focus the discussion of previous works on those similar in terms of interface
aspirations or methodologies.
\newhl{The main differences with our work are: our end-to-end,
design-to-assembly system; the fabrication-aware tight interactive design loop,
and interactive guided assembly plan. To our knowledge, no such complete system
exists.}

\subsection{Design and assembly}
The human-computer interaction and computer graphics communities have embraced
computational fabrication and its evolution beyond classic computed-aided design
and manufacturing (see, e.g.,
\cite{Mueller2015LFS,Umetani2015CTP,Benes2017,Livesu2017MPO}).
\newhl{We join this field of research in rejecting the idea that the existence
of mass-production should preclude an individual's opportunity to participate in
the unique design and customization of everyday objects (e.g., see
\reffig{teaser}).}

We are especially interested in hybrid or heterogeneous systems that combine
3D-printing with other materials to create larger objects.
For example, Kovacs et al.\ \cite{Kovacs2017} build room- and
architectural-scale objects with 3D-printed joints and recycled PET bottles.
%
%
%
\newhl{While their 3D-printed geometry construction is also automatic,
their trusses result from intersecting a 3D shape with a tetrahedral
honeycomb, so joints are less general with fixed topology.}
Kovacs et al.\ \cite{Kovacs2018TPL} incorporate actuation to create articulated
structures, but the fixed joint configuration remains.
In contrast to our design-to-assembly system, the interface contributions of
these methods stop at fabrication: the user is left to build a complex structure
with many labeled parts and no explicit instructions.
\newhl{Unlike this and other tools that only focus on design and fabrication, we
consider the end-to-end system from design to assembly.}

Leen et al.\ \cite{Leen2017SLC} introduce a tangible, modular magnet-based
interface for designing wireframe objects. This work complements ours and could
provide input to our RSDesigner tool, although their sensor rods have upper and
lower bounds on length and joints can only accommodated a fixed number of
incident rods at bounded angles.
Meanwhile, Agrawal et al.\ \cite{Agrawal2015PPS} physically sketch very general,
yet temporary 3D wireframe structures using a device that extrudes tubes of
tape. 

Mueller et al.\ \cite{Mueller2014WPP} break away from layer-by-layer 3D printing
to fabricate wireframe structures by extruding plastic in 3D.
Wu et al. \cite{Wu2016PAM} and Huang et al.\ \cite{Huang2016FRF} extend this
idea to a larger class of wireframe surfaces using a 5DOF printer, while Huang
et al.\ \cite{Huang2018} plan paths for wireframe prints.
Peng et al.\ consider the design of such wire-print objects via a traditional
virtual surface modeling tool \cite{Peng2016OPI} and later an augmented reality
3D drawing interface \cite{Peng2018RIF}.
These methods focus on wireframe representations of \emph{surfaces} and the
design constraints are largely governed by clearance around the printhead during
toolpathing and structural concerns. No assembly is necessary, but structures
are smaller and denser.
%
%

\begin{figure}
\includegraphics[width=\linewidth]{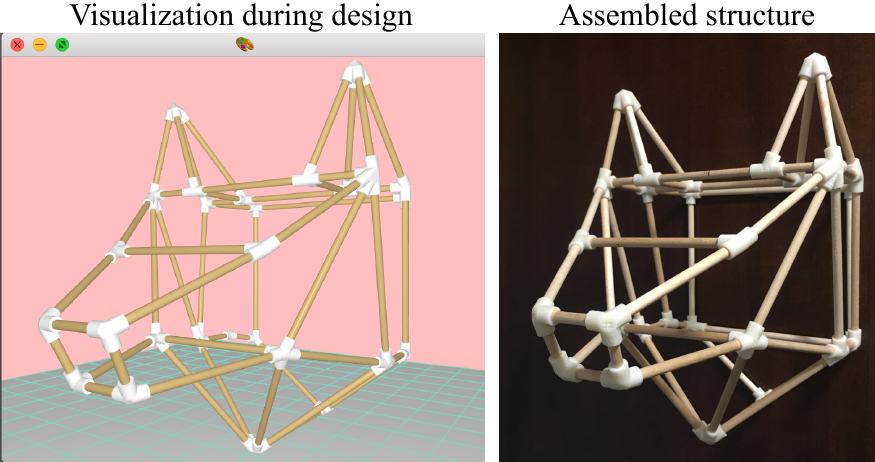}
  \caption{\label{fig:wolf-head}
  RSDesigner displays a visualization closely matching the eventual
  fabrication, reducing opportunity for surprise at assembly time.}
  \vspace*{-0.2cm}
\end{figure}

Recently, Chidambaram et al. \cite{Chidambaram19} introduce a design tool for
wireframe objects constructed via 3D-printed connectors and metal wires. While
their tool provides design suggestions, their method does not detect infeasible
designs due to overlapping rods \newhl{and does not alert the user if their
design will balance.}
\newhl{Their tool computes a stress visualization, but neither complete
description of the method nor accuracy validations are provided. In this design
space, the (strong) wood undergoes bending and stress concentrates at the
(significantly weaker) plastic joints. It is unclear whether the space frames of
Chidambaram et al. are the appropriate model.}
Their method is also restricted both by a hard bound on the length of wires
(3cm) and the angle between rods (35°) in order to safely construct 3D-printed
connectors by unioning sphere and cylinder geometries.
\newhl{
  Due to this strict minimum angle constraint, it would be impossible for their system to
  accommodate the designs in Figures \ref{fig:teaser}~(16°),
  \ref{fig:wolf-head}~(26°), \ref{fig:tesseract-lamp}~(22°), or
  \ref{fig:chip-stand}~(30°).
}
Instead, we propose a more general joint construction algorithm that
accommodates arbitrary angles, sizes, thicknesses, tolerances and polygonal rod
profiles.
\newhl{As a result, our design space is larger and less constraining to the
user.}
Chidambaram et al. provide assembly guidance, but only in the form of
connector/wire indices and a printed lookup table of rod lengths. Our
RSAssembler interface suggests an assembly ordering guided by a focus+context
visualization.

Dritsas et al.\ \cite{Dritsas2017ta} 
create a sequence of \textsc{Grasshopper} scripts to aid in the design of
structures similar to our results.
Given a desired rod diameter, they determine minimum angle of incident joints
allowed by their scripts and prevent/reject designs that do not meet this
criteria. The generated joints are not guaranteed to be solid models which may
cause printer failures.
%
%
%
The interactive design or assisted assembly problems are not considered, so the
user must (presumably) assemble a collection of similar looking parts.

Magrisso et al.\ \cite{Magrisso:2018:DJH} propose a user-assisted process to
generate 3D-printable carpentry joinery.
\newhl{Their goal is different from ours. They seek to enhance traditional
manual carpentry with advanced manufacturing of individual joints, without
placing a strong emphasis on real-time feedback of a tight design loop for the
overall object.}
This process creates intricate joineries. 
\newhl{The design remains creative, but also relies on the user for non-creative
tasks such as supervision of the heuristic when it fails and tuning parameters
to recover a feasible design.}
Our, in comparison, modest joint generation is fully automatic. \newhl{This
allows the user to focus on the creative task of designing the overall object,
facilitated by immediate feedback and accurate visualization.}
The user never concerned with the precise meshing or representation of the joint
geometry, only the high-level design of the structure.
Tian et al.\ \cite{Tian2018MWT} create a library of CNC-millable joineries to
create an interface for woodworking. These beautiful results utilize a different
and complementary fabrication process and design space.

We are inspired by the early interactive exploration work of Umetani et al.
\cite{UmetaniIM12}.
Our contributions are complementary: their method considers loads on panel-based
furniture, but does not consider intersections that would prevent construction
during design exploration.
Later, Garg et al. \cite{Garg2016CDOR} visualize collisions during choreography
and arrangement of space-time reconfigurables, but do not consider geometric
modeling.

\begin{figure*}
  \includegraphics[width=\linewidth]{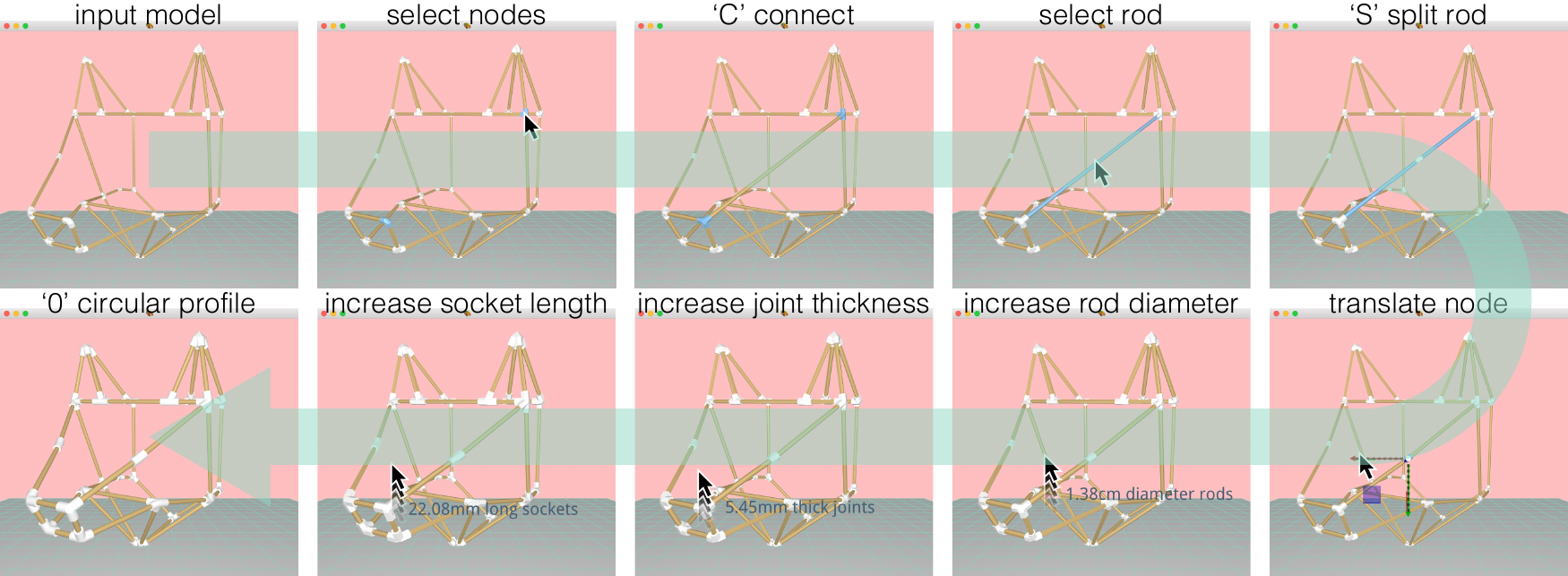}
  \caption{\label{fig:wolf-head-designer-sequence} The user of our design tool
  may conduct a variety of direct manipulation mouse-based editing operations
  and hotkey commands. \newhl{Manipulated values of continuous parameters appear
  on screen directly next to the draggin cursor.}}
  \vspace*{-0.2cm}
\end{figure*}

On a larger scale than ours, Yoshida et al. \cite{Yoshida2015} propose a design
tool and additive manufacturing process to construct architecture-scale
structures out of unstructured chopsticks and glue.
At this scale fused rods behave as a 3D
texture or homogenized material for the shell of the structure.
In contrast, we focus on designs where the rods dominate both the structure's
form and function.

Our design-to-assembly system shares common high-level goals as
\cite{AnnettGWF15,Harquail2016FTI}, who consider the guided design and assembly
of pop-up books and dynamic papercraft objects.

\subsection{Joint Geometry Construction}
It is tempting to consider joint geometry generation as an instance of simple
wireframe meshing, but standard and research methods for this subtly different
problem do not apply to our scenario.
For example, \textsc{Blender}'s Wiremesh Modifier is guaranteed to generate
quadrilateral meshes which is convenient for Catmull-Clark subdivision and other
post-processing, but this method only takes as input edges of a \emph{surface}
mesh.
Panotopoulou et al.\ \cite{Panotopoulou2019} extend this idea to arbitrary
edge-networks by connecting
together variable diameter quadrilateral meshes along each input edge.
Their method minimizes but does not remove the \emph{twisting} of the mesh faces
along the segment.
Unfortunately, any amount of twist is problematic for non-circular profile rods
(see \reffig{rod-orientation-consistency}).

Tonelli et al.\ \cite{TonelliPCS16} create structures from 3D-printed
joints and wooden rods. Their process is not fully automatic
and they only consider the wireframe of a surface mesh
specifically designed to avoid acute angles between edges.
From visual inspection, the method is unlikely to generalize.
Assembly is even more tedious without a guide like RSAssembler: the
joints and rods have been implicitly optimized to have slightly different
geometry.
Their main example took roughly two days to assemble.


Many examples of 3D-printed joints and connectors for furniture-scale
structures can be found online. For example, Gell\'{e}rt \cite{gellert2015}
has gathered a library of
modular 3D-printed connectors for panels to create shelving. Cegar \cite{cegar2014}
constructs 3D-printed joints to connect wooden rods at 0° and 90° angles.  The
startup DesignLibero has a series of furniture and light fixtures composed of
wooden rods and (presumably custom-designed) 3D-printed joints
\cite{rutilo2018}. Fried \cite{Fried2016} has posted a \textsc{Grasshopper} script to
generate node geometry for connecting (presumably only) circular profile rods.
We are inspired by these designs and hope that our reproducible technical
description of joint geometry construction as well as our novel user interfaces
encourage this direction of hybrid design.

Hart's wiremesh generation method \cite{Hart2006} (e.g., as implemented in
\textsc{PyMesh} \cite{pymesh} or \textsc{libigl} \cite{libigl})
provides the foundation for our method.
We identify and correct a few flaws in this method and then extend it to
generate solid geometry compatible for building solid and consistent joints.

%



\section{Fabrication-Aware Design Interface}
\label{sec:designer}
Our investigation is driven by the goal of facilitating the design of rod-joint
structures.
Rod-joint structures afford a harmonious division of complexity. Complex
geometry is delegated to the joints, fabricated by a 3D printer designed for
such a task, while rods retain their intrinsic strength and require only
perpendicular cuts.
Introducing precision 3D-printing into the design task significantly increases
the development time: printing the joints for the guitar stand in
\reffig{teaser} required 10 hours and 10 more hours to remove dissolvable
supports.
We would like to reduce the reliance on time-consuming 3D-printing during design
as well as reduce the probability of fabrication, structural, or assembly
failure.
To this end, we introduce a minimal set of virtual design tools. 
Discarding potential but unnecessary tools is just as important as retaining the
most effective ones.
For this reason, we have written our design tool as a stand-alone application
rather than a plugin to a monolithic commercial CAD tool.
\newhl{For example, existing CAD tools do not deal with intersections well
\cite{Garg2016CDOR}; some tools will simply crash and others will throw an
error.}

The invariant we will maintain in our design tool is a 3D rendering of the
resulting rod-joint structure (see \reffig{wolf-head}).
Three-dimensional joint geometries are rendered in white (i.e., 3D-printed
plastic) and rod geometries in brown (i.e., wood).
We expose the following editing operations to the user:
\begin{itemize}
  \item translate, rotate, and scale selected node positions using a
    standard 3D manipulation
    widget,
  \item connect selected joints with new rods, 
  \item split a selected rod by inserting a joint at the midpoint,
  \item drag on any rod to directly manipulate rod diameter~($2r$), 
  \item drag on any joint to directly manipulate the joint wall
    thickness~($σ$) or the socket length~($h$),
  and
\item choose the number of sides of the polygonal rod profile (or choosing a
  circular profile).
\end{itemize}
See \reffig{wolf-head-designer-sequence} and our accompanying video for
interaction sessions demonstrating each editing interaction.
After any edit, the joint and rod geometries are immediately updated.
When manipulating the sizing parameters ($r$,$σ$,$h$), the new value is
displayed next to the cursor during mouse dragging.

%

\begin{figure}
  \includegraphics[width=\linewidth]{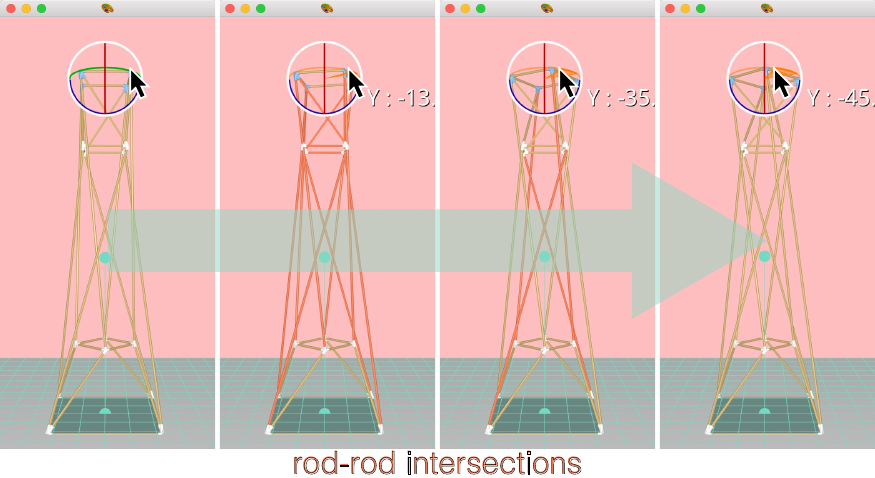}
  \caption{
    \label{fig:tesseract-rod-rod}
    RSDesigner allows the user to manipulate nodes \emph{through} infeasible
    designs, highlighting issues (e.g., rod-rod intersections) interactively so
    the user can creatively explore toward a fabricable design. Real-time
    interaction is key.}
\end{figure}

\subsection{Detecting and highlighting problems}
Not all edge-networks and parameter combinations are fabricable.
We introduce a set of tools to help the user detect potential problems during
virtual design before wasting time trying to fabricate an impossible design.
In the physical world, two rods cannot occupy the same space. In
\refsec{geometry}, we will carefully construct joint geometry to prevent such
rod-rod intersections from happening \emph{locally} at joints. Rod-rod
intersections can also occur \emph{globally} between rods that do not share any
joints. 
We robustly detect rod-rod intersections using the \textsc{libigl} geometry
processing library \cite{libigl} and 
immediately highlight problematic rods in red.
We do not \emph{prevent} the user from making invalid designs. It is often
desirable to traverse through invalid states into a new valid state (see
\reffig{tesseract-rod-rod}). Our real-time feedback allows the user to visually
track the feasibility of the design during any edit.

\begin{figure}
  \includegraphics[width=\linewidth]{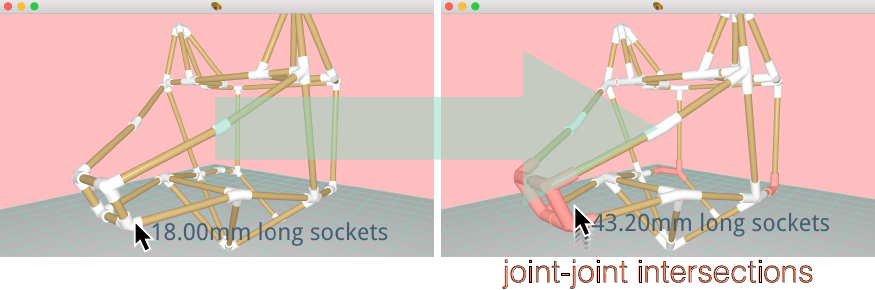}
  \caption{
    \label{fig:joint-swallowing} 
    The user drags on the joint to increase the socket length. If the joints
    become too large and intersect, RSDesigner highlights them to alert the user
    of an infeasible design.}
  \vspace*{0.5em}
  \includegraphics[width=\linewidth]{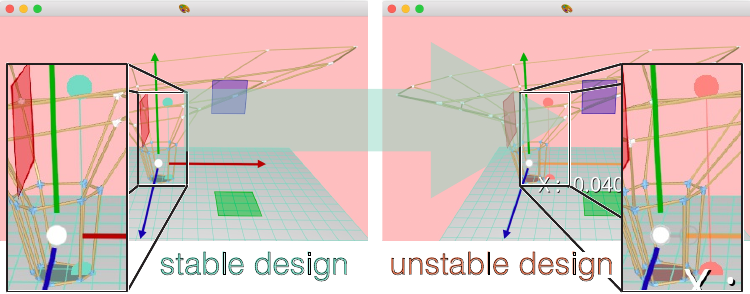}
  \caption{
    \label{fig:pavilion}
    The user translates the base of this pavilion slightly to the left and
    RSDesigner highlights that the design will no longer balance, by coloring
    the center of mass and support polygons red.
  }
  \vspace*{-0.5em}
\end{figure}

The angles of rods incident on a joint and the rod/joint thickness parameters
determine the ultimate geometry of a joint. If the joints become too large or
the rods between them too small, the joint geometries will overlap,
\emph{swallowing} a rod (in the notation of the next section, if
$g_{ij}+g_{ji}+2h>\ell_{\{i,j\}}$). We immediately highlight such problematic
joints in red, alerting the user of an inefficient or undesirable design (see
\reffig{joint-swallowing}).

We also help the user determine whether the current design will stand.  If the
center of mass projected onto the ground falls outside of the support polygon,
the design is deemed unstable (see, e.g., \cite{PrevostMIS2013}), and we alert
the user by highlighting the center of mass and support polygon red (see
\reffig{pavilion}).

The effectiveness of our design tool hinges on the ability to efficiently and
fully automatically generate general and fabricable joint geometries and rod
lengths. We now turn our attention to constructing and then fabricating these
geometries.

\begin{figure*}
\centering
  \includegraphics[width=\linewidth]{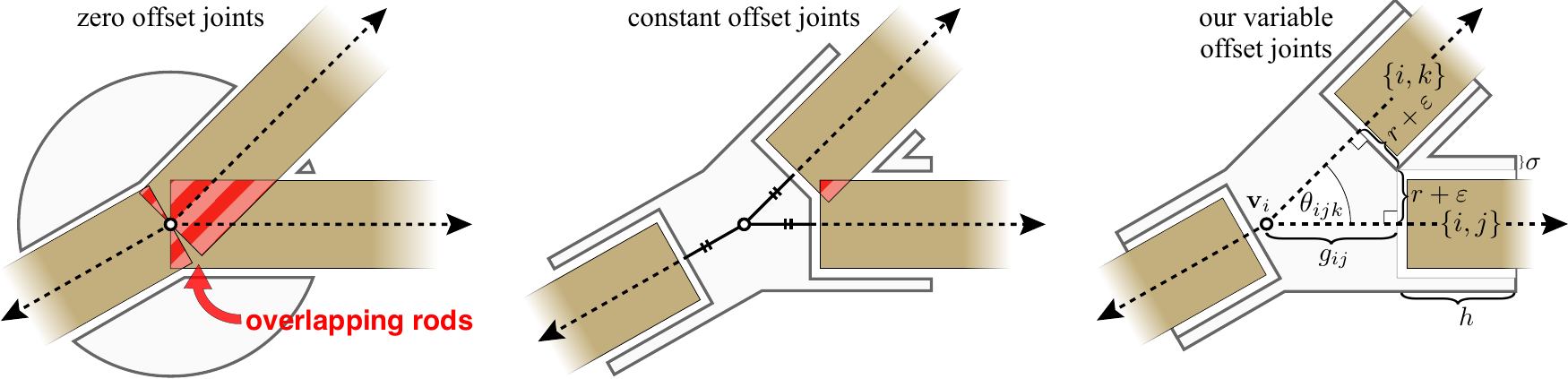}
  \caption{
    \label{fig:offset-comparison}
    Previous joint geometry generation methods do not consider the possibility
    of rod intersections at joints. A per-edge offset is necessary and can be
    minimized on a case-by-case basis.
  }
\end{figure*}

\section{Geometry \& Fabrication}
\label{sec:geometry}

The input to our geometry construction algorithm is a 3D edge-network, i.e., a
graph embedded in $\R³$, composed of a list of $n$ node positions $\vv_i∈\R³$
for $i∈\{1,…,n\}$ and a list of $m$ \emph{undirected} edges as pairs $\{i,j\} ∈
\{1,…,n\}²$ where we use the equivalence relation $\{i,j\} = \{j,i\}$. 

In this technical section, we use subscript notation such that
$a\ijs = a_{\{j,i\}}$ refers to a quantity associated with the undirected
edge $\{i,j\}$, whereas $b_{ij}$  refers to a quantity associated with
end-point $i$ on the edge $\{i,j\}$ and in general $b_{ij} ≠ b_{ji}$. In most
cases, the difference will also be clear from context.

The algorithm is controlled by a number of user-defined parameters (see
\reffig{offset-comparison}, right):
$r$ the radius of the rods, $p$ the number of sides on the polygonal
cross-sectional profile of the rods (without loss of generality, we will assume
these polygons are regularly shaped), $σ$ the thickness of the 3D-printed
\emph{joints} encasing each node, $h$ the amount that joints overhang along
incident rods, and $ε$ the ``engineering tolerance'' (possibly negative for
friction fitting) between the joints and rods.

The output of our method includes $n$ solid meshes representing the
surfaces of the 3D-printable
joint geometry at each node and $m$ precise rod lengths to cut.
As seen in \reffig{offset-comparison}, the physically realizable length of the
rod of each edge $\{i,j\}$ will generally be less than the raw edge length
($‖\vv_i - \vv_j‖$). Instead, the precise lengths will be implicitly determined
by the geometry of the joints at either node.

\begin{figure}
\centering
\includegraphics[width=\linewidth]{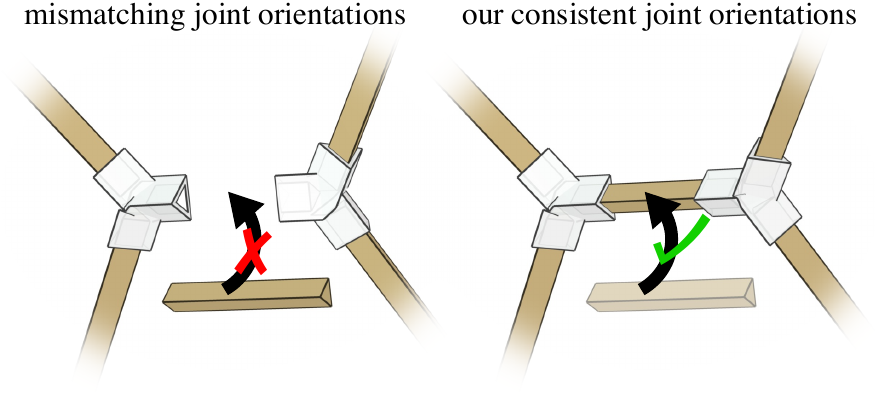}
\caption{\label{fig:rod-orientation-consistency}
  Direct extensions of wiremeshing algorithms (e.g.,
  \protect\cite{Panotopoulou2019,Hart2006}) may result in \emph{twisting} along edges,
  leading to inconsistent rod-orientations at either end. Our joint-construction
  algorithm avoids this.
  }
\end{figure}

\subsection{3D-Printable Solid Joint Geometry}

The geometry of the joint at each node will be an independent solid object,
but we require that the outlets at either joint incident on an edge to be
\emph{consistent} so that polygonal-cross-section rod geometry can be
rotated to fit either end \reffig{rod-orientation-consistency}.

A useful subroutine is to generate the primitive geometry of a solid mesh of a
generalized cylinder with the profile of a $p$-sided polygon. This is
accomplished by extruding a regular $p$-gon inscribed in the unit circle of the
$xy$ plane along the $z$-axis for one unit.

This unit-cylinder mesh geometry can then be transformed to lie along any given
edge.
For each edge $\{i,j\}$, we compute a 3D rotation $\RR\ijs ∈ SO(3)$ aligning
the $z$-axis vector $\e_z = (0,0,1)^\top$ to its unit edge vector
$\hat{\w}_{ij} =
(\vv_j-\vv_i)^\top/‖\vv_j-\vv_i‖$:
\begin{equation}
\label{equ:rotation-between-two-vectors}
  \RR\ijs  = \I + [\e_z×\hat{\w}_{ij}]_× + \frac{1}{1+\e_z⋅\hat{\w}_{ij}}[\e_z×\hat{\w}_{ij}]_×² 
\end{equation}
where $\I∈\R^{3×3}$ is the identity matrix and $[\x]_×$ is the 
matrix representing cross-product by the vector $\x$:
\begin{equation}
  [\x]_× = \left(\begin{array}{ccc}
    0 & -x_3 & x_2 \\
    x_3 & 0  & -x_1 \\
    -x_2 & x_1 & 0
  \end{array}\right).
\end{equation}

We \emph{could} place rod geometry along each edge $\{i,j\}$
by composing this per-edge rotation with anisotropic pre-scaling along the
$z$-axis by the edge length $‖\vv_j-\vv_i‖$ and radially in the
$xy$-plane by the desired radius $r$ and a post-translation to
the edge tip position $\vv_i$. As a per-edge affine
transformation:
\begin{equation}%
\begingroup%
\setlength{\arraycolsep}{2pt}%
  \underbrace{\left(\begin{array}{cc}
    \I &
    \begin{array}{c}
      \vv_i
    \end{array}\\
    0\ 0\ 0 & 1
  \end{array}\right)
  \left(\begin{array}{cc}
    \RR\ijs  &
    \begin{array}{c} 0\\ 0\\ 0 \end{array}\\
    0\ 0\ 0 & 1
  \end{array}\right)
  }_{\T\ijs }
  \left(\begin{array}{cccc}
      r&0& 0&0\\
      0&r& 0&0\\
      0&0& ‖\vv_j-\vv_i‖&0\\
      0&0& 0&1
  \end{array}\right),
\endgroup%
\end{equation}
where $\T\ijs  ∈ \R^{4×4}$ is a rigid transformation \emph{placing} the rod into
the edge-network. However, this naïve rod geometry will result in messy intersections at each node
(see \reffig{offset-comparison}, left).
In the physical world, we can not allow the rods of multiple edges incident on a
node to ``share'' the space at that node.
Offsetting by a uniform amount $g$ (cf.\ \cite{Hart2006}) works when $g$ is
large relative to $r$ and the angles between incident rods are not very acute.
The 3D-printed joints of Tonelli et al.\ \cite{TonelliPCS16} use a fixed offset,
but their results are limited to surface edge-networks with modest angles.
For arbitrary edge-networks, acute angles are common.
%
If $g$ is too small relative to $r$, overlaps will occur even for obtuse angles
(see \reffig{offset-comparison}, center).
If $g$ is too large, joints become bulky (everywhere) and may even
(unnecessarily) envelope small edges.

\begin{figure*}[b!]
  \includegraphics[width=\linewidth]{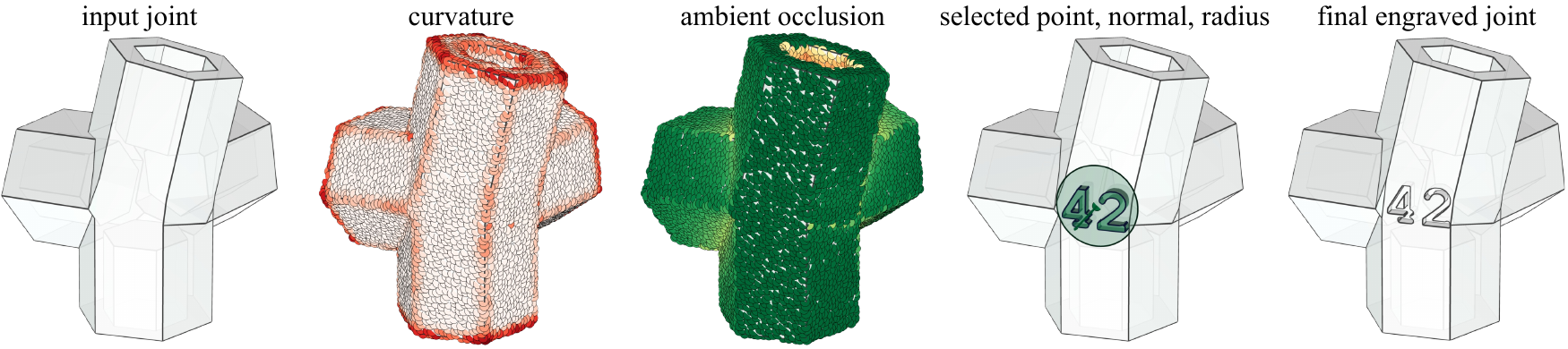}
  \caption{\label{fig:engraving} We introduce a subroutine to geometrically
  engrave ids into each joint part using curvature and occlusion to find a
  readable location.
  }
  \vspace*{-0.5em}
\end{figure*}

We \emph{could} remove the rod-intersection volumes (e.g., the red regions in
\reffig{offset-comparison}) from the rod geometries, but this would require
non-trivial shaving or whittling of the rods.
Instead, we move all complex geometry to the 3D-printed joints and use simple
straight perpendicular cuts on off-the-shelf rods.

To this end, we compute per-node-edge offsets, where $g_{ij}$ is the offset at
node $i$ along the incident edge $\{i,j\}$. The offsets at either end of an edge
$\{i,j\}$ will in general be different (i.e., $g_{ij} ≠ g_{ji}$).
We would like $g_{ij}$ values that: 1) are as small as possible and 2) guarantee
that rods will not overlap.
We can compute a safe offset $g_{ij}$ by considering
the minimum angle formed by edge $\{i,j\}$ and all other edges $\{i,k\}$ with
$k≠j$:
\begin{align}
  θ_{ij} &= \min_k \cos^{-1} \left( \hat{\w}_{ij} ⋅ \hat{\w}_{ik} \right) \\
  \intertext{or equivalently the largest dot-product}
  c_{ij} &= \cos θ_{ij}  = \max_k\ \hat{\w}_{ij} ⋅ \hat{\w}_{ik}.
\end{align}
In general, for a node $i$ the smallest angles along different incident edges
will not be the same (i.e., $θ_{ij} ≠ θ_{ik}$).
Given the rod radius $r$ and engineering tolerance $ε$, a safe offset $g_{ij}$
is the solution to a trigonometry problem solved using the tangent half-angle
formula:
\begin{equation}
  g_{ij} = (r+ε) \sqrt{\frac{1+c_{ij}}{1-c_{ij}}}.
\end{equation}
As this formula confirms, the offset tends toward infinity as the angle tends
toward zero (and $c_{ij}$ tends toward one).


Armed with offsets that guarantee the absence of rod intersections at joints, we
can now generate solid joint geometry.
We start by considering every edge $\{i,j\}$. We generate unit-cylinder mesh and
scale it radially by $r+σ$ and axially by $h+σ$.
We then place two copies offset axially by
$g_{ij}-σ$ and $‖\vv_j-\vv_i‖-h-g_{ji}$, respectively. Both are finally
transformed into place by $\T\ijs $.
All together, the \emph{tip} and \emph{tail} pieces are transformed,
respectively, by:
\begin{align}
  \mat{H}_{ij} &= \T\ijs  
  \left(\begin{array}{cccc}
      r+σ&0& 0  &0\\
      0&r+σ& 0  &0\\
      0&0&   h+σ&g_{ij}-σ\\
      0&0&   0  &1
  \end{array}\right) \\
  \intertext{and}
  \mat{H}_{ji} &=
  \T\ijs  
  \left(\begin{array}{cccc}
      r+σ&0& 0&0\\
      0&r+σ& 0&0\\
      0&0&   h+σ& ‖\vv_j-\vv_i‖-h-g_{ji}\\
      0&0&   0&1
  \end{array}\right).
\end{align}
We denote the transformed solid models as $\C_{ij}$ and $\C_{ji}$
respectively.
Though strictly not necessary to generate a solid joint, the $+σ$ in the axial
scaling ensures a $σ$-thick ``cap'' at each end of a rod.
For each pair of transformed cylinders, we keep track of the mesh vertices of
the cylinder model that end up at either end of the edge. That is, those with
projected distance $g_{ij}$ and $g_{ji}$ to the tip and tail nodes,
respectively. We call these vertex-sets $\V_{ij}$ and $\V_{ji}$, respectively.
Due to the procedural generation and transformation of the cylinder model, this
bookkeeping is purely combinatorial and does not require measuring distances
\emph{after} transforming each cylinder model.

Now we consider each node of the input edge-network. Like Hart's method
\cite{Hart2006}, we compute the convex hull $\H_i$ of the node position $\vv_i$
and all vertex-sets $\V_{ij}$ from incident edges $\{i,j\}$. While this convex
hull is guaranteed to have \emph{at least} a two-dimensional intersection with
each of the incident cylinder models, it is not true (cf. \cite{Hart2006}) that
faces of the cylinder meshes will always appear as faces of the convex hull:
some vertices of $\V_{ij}$ may be strictly \emph{inside} the
convex hull.
We merge the hull model $\H_i$ and the transformed cylinders $\C_{ij}$ of each
incident edge $\{i,j\}$ by computing their exact mesh union via
\cite{Zhou:2016:MASG}. We denote the solid result of this union~$\U_i$.
%

As a side effect of this process, we have determined that the precise length
$\ell_{ij}$ to cut the rod at each edge $\{i,j\}$ is the full edge-length minus
the safe offsets computed at either end:
\begin{equation}
  \ell_{ij} = ‖\vv_j - \vv_i‖ - g_{ij} - g_{ji}.
\end{equation}
For each edge $\{i,j\}$, we generate unit edge-cylinder
mesh and scale it radially by $r+ε$ and axially
by $\ell_{ij}+2ε$.
We then align this geometry with the offsets and transform it into place by
by $\T\ijs $.
All together, the transformation of the unit-cylinder per-edge is:
\begin{equation}
  \T\ijs  
  \left(\begin{array}{cccc}
      r+ε&0& 0  &0\\
      0&r+ε& 0  &0\\
      0&0&   \ell_{ij}+2ε&g_{ij}-ε\\
      0&0&   0  &1
  \end{array}\right).
\end{equation}
We denote the transformed model at each edge by $\E\ijs$.

To complete our joint geometry, we consider each node again. We compute the
exact solid difference 
of the joint geometry $\U_i$ and the geometry $\E\ijs$ of all incident edges
$\{i,j\}$. The result of this Boolean operation is the final solid joint
geometry
\begin{equation}
  \J_i = \U_i \setminus \bigcup\ijs \E_{ij}.
\end{equation}

The mesh boolean operations resulting in $\U_i$ and then $\J_i$ are necessary to
create a solid mesh.
During interaction with our fabrication-aware interface, the user is unaware
that boolean operations are happily skipped as they do not affect the visual
appearance.

\subsection{3D Printing}
Joint geometries $\J_i$ are 3D printed using heuristic to pick a printing
direction that minimizes support material placed \emph{inside} the cavities at
each socket (the most difficult to dissolve/remove).
We compute the rotation that aligns the average edge-vector $\bar{\w}_i = ∑\ijs
\hat{\w}_{ij}$ with the printing extrusion direction (similar to
\refequ{rotation-between-two-vectors}).
We use existing software (\textsc{GrabCADPrint} or \textsc{Simplify3D}) to pack
the rotated 3D geometries into the smallest number of build volumes.

Each joint is automatically engraved with a two-digit identification number.
This is achieved fully automatically. We start by oversampling the joint
geometry at 10,000 uniformly random locations. We estimate curvature at each
sample by taking the distance-weighted average of dihedral angles between the
$k=200$ nearest neighbors. We compute ambient occlusion at each sample with
respect to the original geometry. We select the sample with the smallest sum of
these two values and set the engraving's radial extent to the distance to
furthest of the sample's $k$ neighbors. Text geometry with a thickness of $σ/2$
is placed accordingly and subtracted from the joint geometry via
\cite{Zhou:2016:MASG} (see \reffig{engraving}). The curvature and occlusion term
encourages the engraving to be centered on a flat and visible region,
respectively. This simple method was surprisingly effective. The label
visibility allows RSAssembler to better guide the assembly described in the next
section.
It would be interesting to extend our labeling and print orientation heuristics
to consider perceptual preference \cite{Zhang2015PMP}.

\paragraph{Rods}
In theory, any cutting method could be used to cut the $m$ rods (e.g., a
traditional hand saw and miter box). However, often each joint has a unique
length $\ell_{ij}$ and the setup/measuring time of each cut starts to dominate
for manual methods when the number of rods $m$ is large. Instead, we use a laser
cutter to precisely cut all rods (or as many will fit into the machine at once)
simultaneously. To maximize the efficiency of this process, we solve a
one-dimensional ``bin-packing''. That is, given a set of $m$ desired lengths to
cut and raw factory uncut rods of factory-length $b$ (e.g., $b=1\text{m}$), we
find the assignment of cuts to uncut rods that minimizes waste and uses the
fewest uncut rods (see \reffig{laser-bin-packing}).
We implemented a ``first-fit'' algorithm with multiple random orderings and
taking the best packing (the optimal fit, while NP-hard to compute, will
correspond to the first-fit result of \emph{some} ordering \cite{Lewis09}).
We use bin sizes of $b-2p$, where $p$ is a padding amount (e.g., $p=1\text{cm}$)
to account for rough ``factory'' cuts at either end of the rod. The optimized
cuts are translated into line segments of a \texttt{.svg} file that is sent to
the laser cutter to be cut in a single run (see \reffig{laser-bin-packing}).

We experimented with using the laser-cutter to mark each rod during cutting in
the hopes that this helps assembly. We found this to be unnecessary and, in
fact, confusing.
%
Instead, it was much faster to simply presort all of the cut rods by length and
have a ruler/calipers nearby to select the desired rod.

%
%

\section{Guided Assembly Interface}
\label{sec:assembler}
After physical fabrication, the design is reduced to a large number of 3D
printed of joints and laser-cut rods.
For complex designs (e.g., the wolf head in \reffig{wolf-head}), there are many
similar looking joints and similar length rods.
Incorrect placement of a rod is often not a problem as this \emph{length} error
will diffuse through the design.
In stark contrast, incorrect placement of a joint leads to an \emph{angle} error
that grows linearly with distance and can cause a Domino-effect of
misalignment, infesting the entire design.
To help the user avoid such assembly disasters, we propose a guided assembly
interface.

The user starts by organizing the 3D-printed joints (e.g., sorted by
engraved id) and laser-cut rods (e.g., sorted by length).
On a nearby screen (e.g., currently our app runs on a MacBook, but could be
ported to a tablet/phone), our guided assembly tool shows a 3D visualization of the
design.
The user can manipulate the camera parameters to view the design from any
direction.
After placing a rod or joint into position, the user hits a hotkey and the
interface proceeds to the next suggested part to place.

\begin{figure}
  \includegraphics[width=\linewidth]{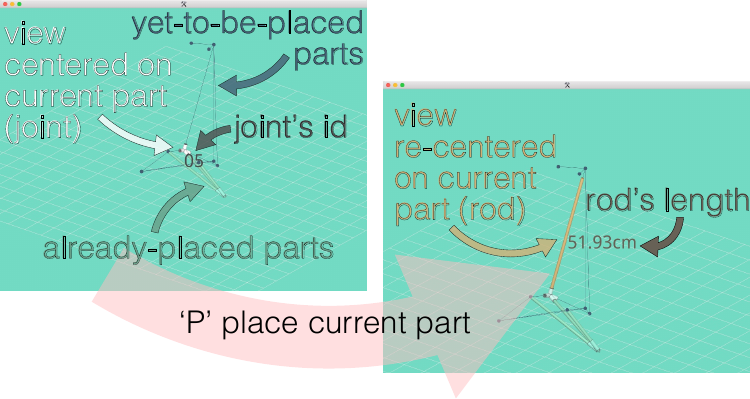}
  \caption{\label{fig:guitar-stand-assembler}
  The RSAssembler interface embraces a focus+context design.
  \newhl{The current object is centered and the zoom is initialized to fit the entire
  object in view in case the user wishes to tumble the camera. (Screenshots are
  intentionally \emph{not} cropped to the object.)}
  }
\end{figure}

\begin{figure}
  \includegraphics[width=\linewidth]{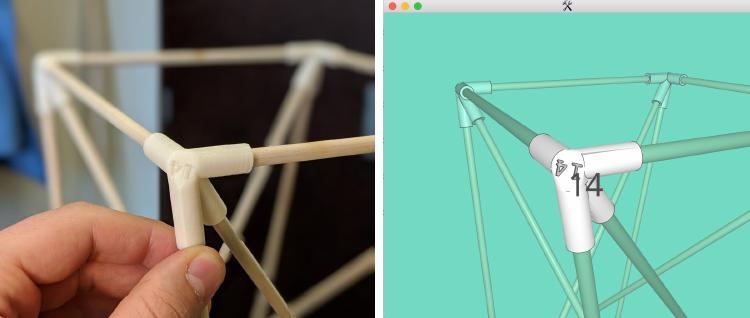}
  \caption{\label{fig:joint-assembler-zoom-in}
  The joint geometries shown by RSAssembler match the 3D printed. Engraved
  joint IDs also facilitate orientation matching.
  }
\end{figure}

Our guided assembly interface exercises focus+context \cite{CockburnKB08} (see
\reffig{guitar-stand-assembler}): the current part to be placed is rendered in
full color matching the physical counterpart (e.g., white for 3D printed joints
and brown for wooden rods).  Already-placed parts are shown to provide context
but \emph{recessed} out of focus by shifting their color toward the (non-white)
background color (e.g., teal). Yet-to-be-placed parts are abstracted as dots
(joints) and line segments (rods). When the user signals that the current part
has been placed, the 3D camera smoothly transitions to focus on the next part.
That part is placed at the center of the screen and viewed at a distance so that
all connected parts will fit into view when rotating the camera. For joints, the
id number is show in a large font and the geometry (with engraving) is shown in
a high-contrast rendering style to assist in matching the correct orientation
(see \reffig{joint-assembler-zoom-in}).
\newhl{For symmetric joints, the engraved ID replicated in the RSAssembler
display serves as a further registration mark.}
For rods, the length is displayed.
See the accompanying video for a full guided assembly sequence.

Random assembly order would require significant context switching, both visually
and physically.
In addition, we found that is much more difficult to merge multiple sub parts
than to add pieces one-by-one.
\newhl{We experimented with various assembly order heuristics breadth-first, rod
length priorities, etc.), and ultimately found that 
a depth-first traversal of the nodes in the edge-network  works best.
This strategy is guaranteed to produce a complete ordering, regardless of the
size of the input.}
After placing a joint, the tool suggests each not-yet-placed incident rod and
then proceeds to the next not-yet-placed adjacent joint.
This ensures that the set of already-placed parts is always connected and that
the user often adds a new joint adjacent to the last added joint.
\newhl{In our experiments, structural stability of the partially assembled
objects was not an issue, though as a future improvement this could be taken
into account in the ordering (or even the possibility of adding temporary
assembly-only rods).}

\section{Results \& Discussion}
We 3D print the joints in our results using a Stratasys F170 in ABS Ivory
plastic with dissolvable QSR support material.
We laser cut our rods using a Trotec Speedy400 Flexx, which has a 1m×0.6m build
plate.
We use a variety of different rod materials purchased from a local hardware
store: 6.35mm and 12.7mm diameter round hardwood dowels and 10.5mm²
square-profile wooden molding.
The bottleneck in our end-to-end design-to-assembly system is by far the
3D-printing and support-material dissolution. 
Our slow 3D-printer took more than 10 hours for most examples (10-20 joints) and
automatic support-material-removing bathing took nearly as long.
Fortunately, the entire 3D-printing process is fully automatic (aside from
moving parts from the printer to the bath).
Using a ``comb'' jig to hold rods in place, cutting takes a few
minutes.
Assembly itself lasted under 45 minutes for all examples included in this paper:
the wolf head in \reffig{wolf-head} taking longest with 28 joints and 52
rods.

We printed a small board containing sample sockets of varying engineering
tolerances $ε$ to determine good values for each rod radius and profile we used.
This way we avoid printing joints that ultimately do not friction fit our rods
(due to inaccuracies of either part).
\newhl{Tight friction fit joints allow us to avoid the use of tools}
\setlength{\columnsep}{10pt}%
\begin{wrapfigure}[14]{r}{1.5in}
  \vspace*{-0.1cm}
\includegraphics[width=\linewidth]{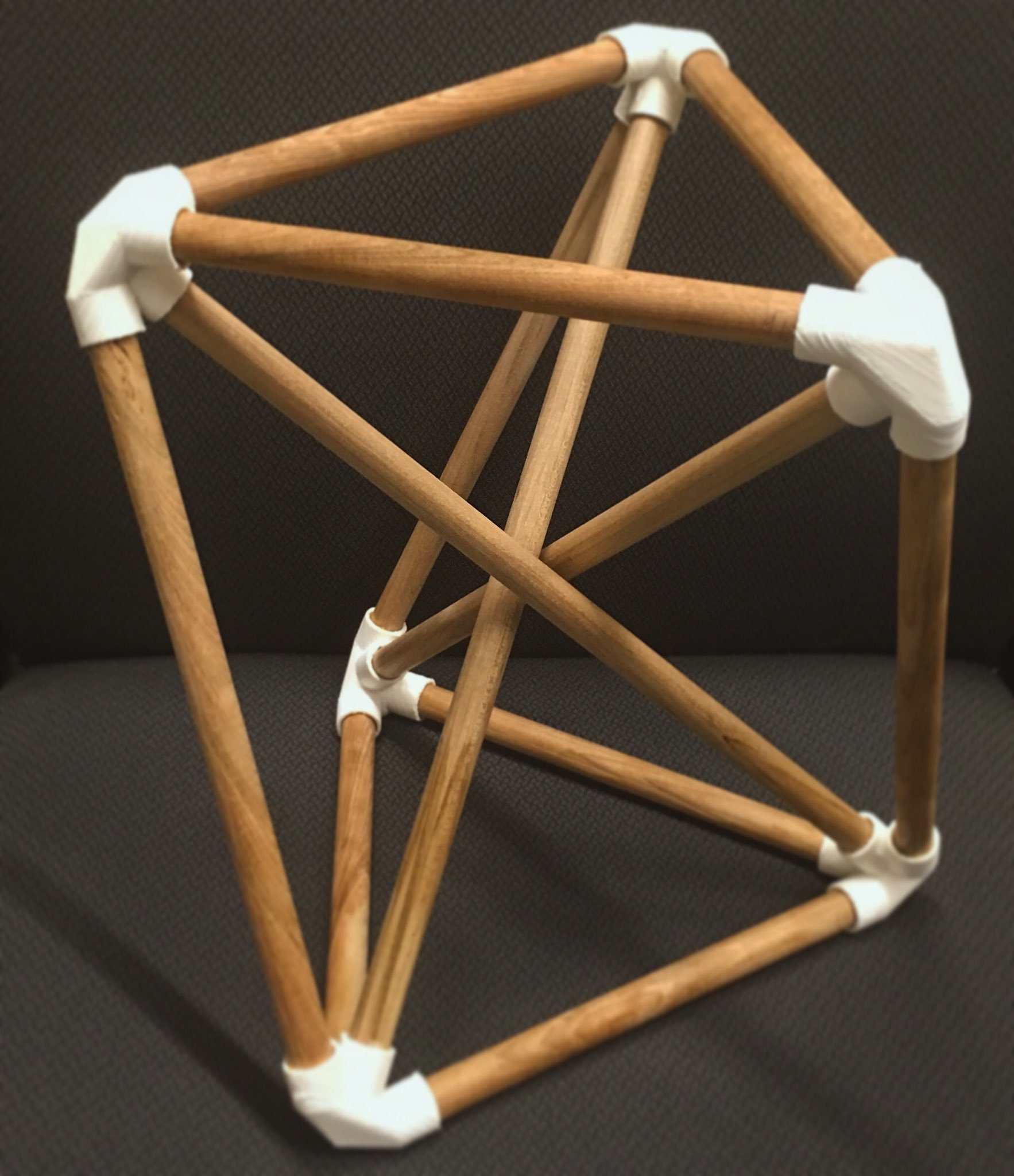}
  \caption{
    \label{fig:little-schoenhardt}
    Nearly touching rods.
  }
\end{wrapfigure}
\newhl{
or screws during assembly.  A locking mechanism and loose-fit joints would be an
interesting alternative.}

The rod-rod intersection testing allows designers to create structures with
closely packed but not intersecting rods. In
\reffig{little-schoenhardt}, the triangular
prism shape was twist just until a collision occurred. Indeed, the assembled
rods are nearly touching. Finding designs like this without a virtual 
real-time feedback interface like RSDesigner would be tedious and difficult.

\begin{figure}
  \includegraphics[width=\linewidth]{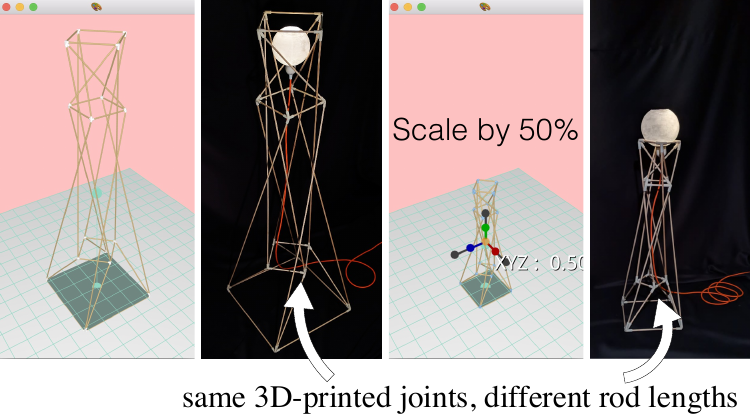}
  \caption{\label{fig:tesseract-lamp} Isotropic scaling the design will not change the joint
  geometry, so the same 3D prints may be reused at different scales.}
\end{figure}

An interesting design feature that serendipitously emerged from exploring in
RSDesigner is that when scaling the node positions isotropically, the joint
geometry remains identical: only the rods shrink/grow uniformly.
This informs us that the same 3D printed joints can be used to create different
scales of the same object: just replace the rods.
We realized this feature in the lamp design in \reffig{tesseract-lamp}. A
1.4m-tall lamp is first constructed, then all rods are cut in half and the lamp
is reassembled as a 0.7m-tall lamp.

\setlength{\columnsep}{10pt}%
\begin{wrapfigure}[15]{r}{1.7in}
  \vspace*{-0.4cm}
\includegraphics[width=\linewidth]{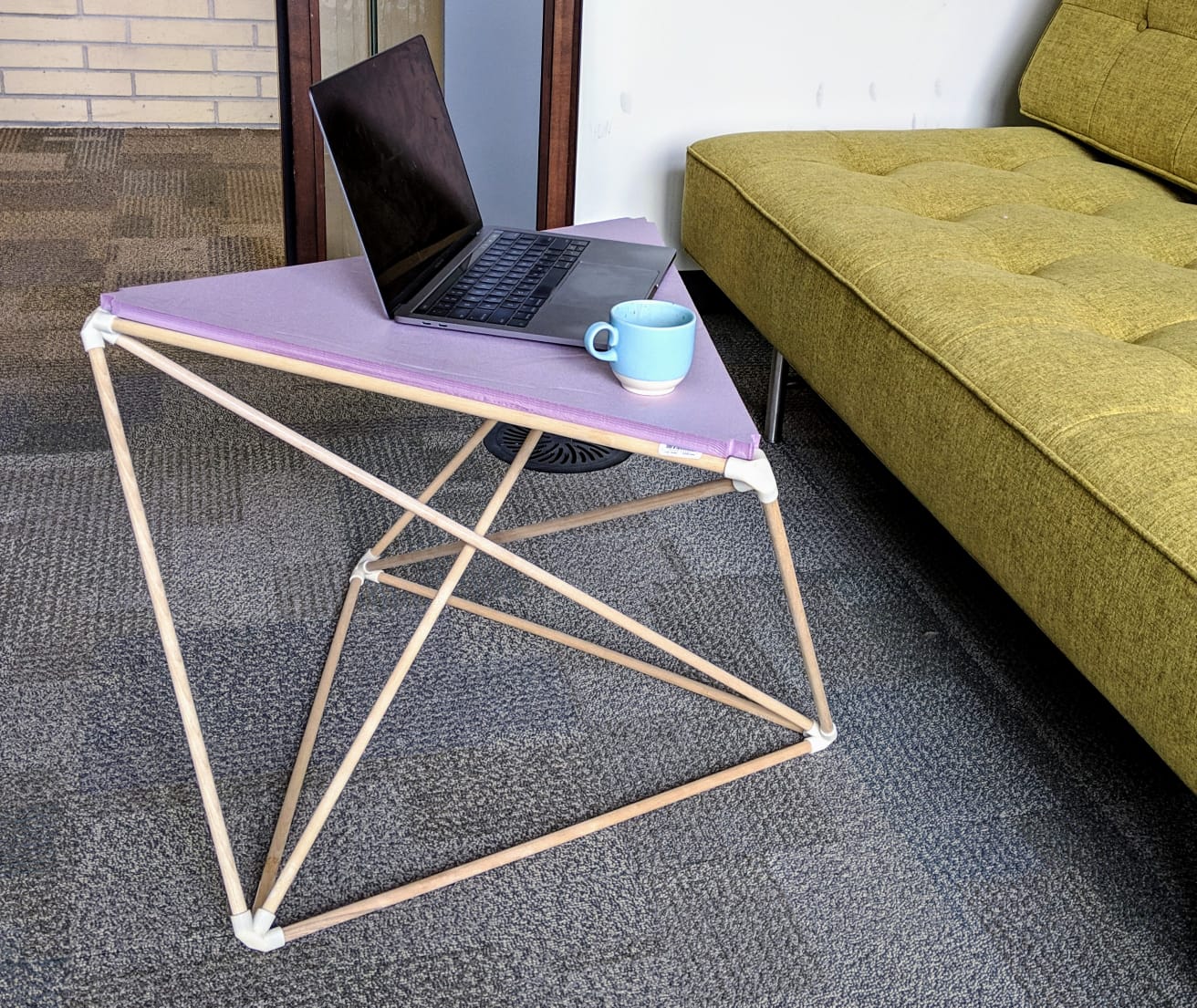}
  \caption{\label{fig:coffee-table} 
  \newhl{
    This tailored coffee table inherits the strength of hardwood dowels to
    supports not only itself, but also top panel and laptop (table in daily use
    for six months).}
  }
\end{wrapfigure}

We do \emph{not} perform finite-element analysis or optimize the design for
structural stability beyond balance.
Nonetheless the structures we create are sturdy, inheriting the strength of the
wooden rods.
In \reffig{coffee-table}, we demonstrate a minimalist coffee-table design.
This table is now in regular use at an office environment.

Disassembled, our structures are very compact (see \reffig{teaser}). Further,
rods can be cut from standard-size dowels. Transporting a design could be as
simple as shipping the joints and cutting rods on site.

\begin{figure}
  \includegraphics[width=\linewidth]{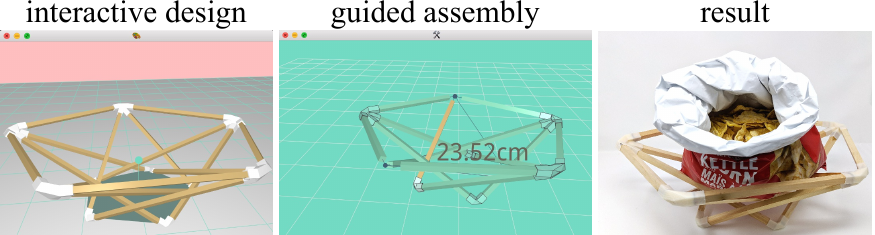}
  \caption{\label{fig:chip-stand} The square-profile of the rods in this
  design affect not just the appearance but also the constructed
  rod geometries.}
\end{figure}

Although circular-profile dowels are the most common and cheapest rods
purchased at a hardware store, our entire system supports polygonal profile
rods. In \reffig{chip-stand}, a decorative bowl is designed using square-profile
molding for rods.

\newhl{Unlike cheap, mass-manufactured guitar stand, the unique design shown in
\reffig{teaser} represents an aesthetic discovered by the user during modeling.
The interface tools and on-the-fly evaluation of RSDesigner allow a user to
fine tune a design in a tight loop.}

\section{Limitations \& Future Work}
Our design-to-assembly workflow is not without limitations.
We focused on mainly furniture-scale results; although our tools will function
correctly at any scale, the size of the structure is bound on the small end by the
precision of the 3D printer and on the end by the length of the largest rod.
We have also left various incremental improvements to design space as future
work: e.g., non-regular convex polygon profiles, varying the rod radius $r$
across edges, varying thickness $σ$ across joints, adding point loads to adjust
the center-of-mass.
We are intrigued by the idea of adding fine-scale decorative patterns
\cite{SchumacherTG16} or Voronoi duals \cite{Magrisso:2018:DJH} to our joint
geometries.
Very large or volume-filling designs (e.g., trusses) might raise performance
concerns and require a dynamic bounding-volume hierarchy (cf.\
\cite{Garg2016CDOR}) for intersection testing etc.
Our assemblies only required a single person guided by RSAssembler.
We forgo a formal user study to confirm that guided assembly is better than
having no assembly instructions \cite{Kovacs2017} or a static plan
\cite{TonelliPCS16}.
It would be interesting to use crowds or robots to build architectural
scale structures \emph{à la} Lafreniere et al.\ \cite{Lafreniere2016CF}.

If we idealize the joints and rods as perfectly rigid objects, then all our
results are mathematically impossible to assemble. We are reliant on compliance.
%
%
%
For our sparse, 1D structures reachability was not a
major concern (cf.\ panel-based structures, \cite{UmetaniIM12}).

The full power of the laser cutter has not truly been leveraged. We use the laser
cutter out of availability (we have one), convenience (easier than manually
cutting), and efficiency (faster, too).
For the results presented here a robotic rod cutter might be more appropriate. 
In future work, it would be interesting to further exploit at least the
two-dimensional capabilities of the laser cutting in conjunction with the
interfaces presented here.

\begingroup
\setlength{\columnsep}{10pt}%
\begin{wrapfigure}[10]{r}{1.7in}
  \vspace*{-0.4cm}
\includegraphics[width=\linewidth]{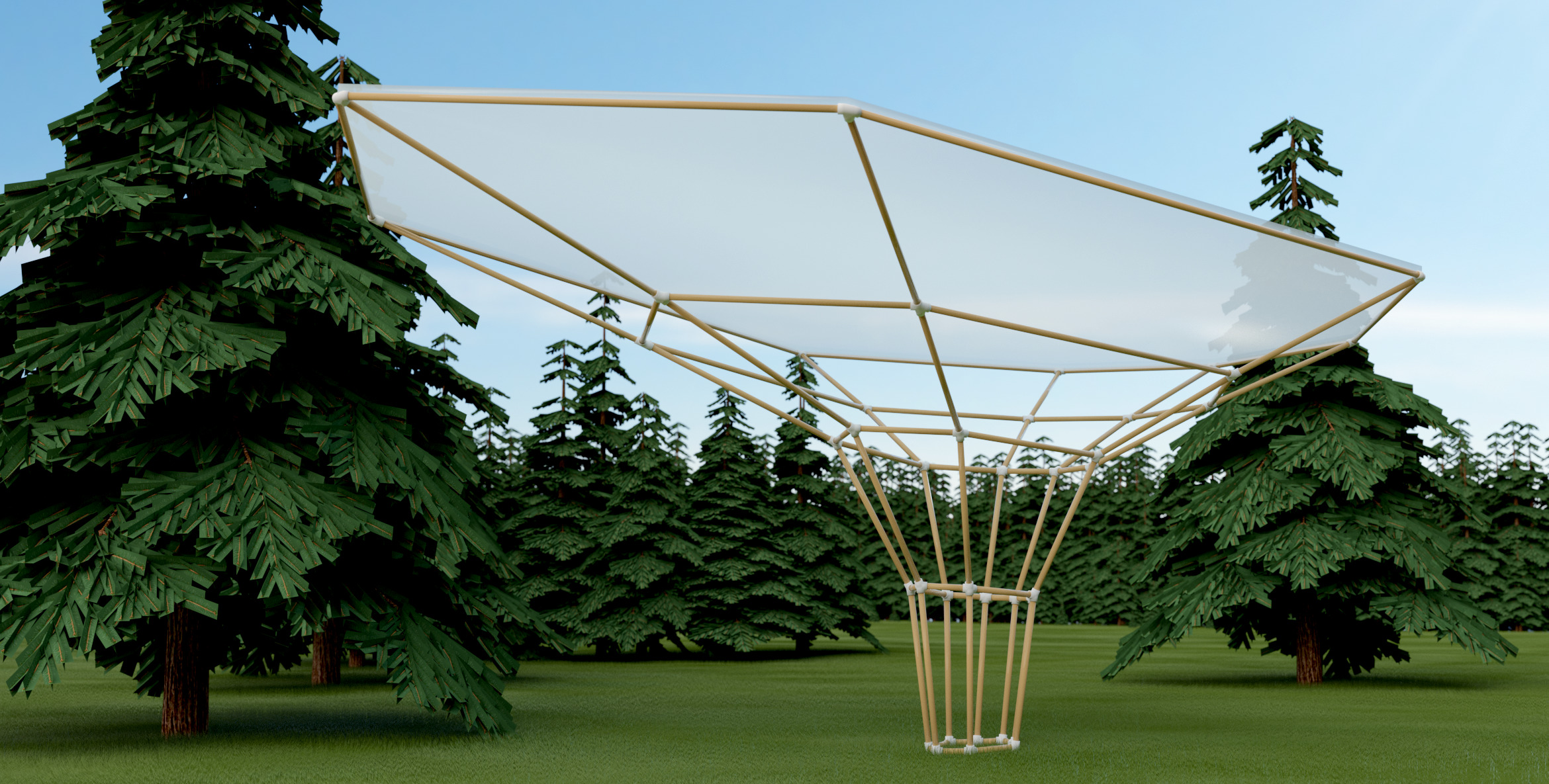}
\caption{\label{fig:pavilion-rendering} RodSteward could extend
  design-to-assembly beyond furniture: a rendered pavilion.}
\end{wrapfigure}
While our lamps and coffee tables can withstand light external loads, larger
furniture or structural shapes would require consideration of its use and
placement its environment \cite{UmetaniIM12,Whiting2017EFR}. In future work, we
would like to consider structural properties of joints
\cite{mueller2014additive} and directly incorporate loads from non-trivial
contact and friction with external objects into our design tool (i.e., beyond
the point loads of \cite{UmetaniIM12}).
We have already entertained interest from architects to adapt our joint
generation for architectural-scale objects (see, e.g.,
\reffig{pavilion-rendering}). We are excited by the recent work of Yoshida et
al. \cite{Yoshida2019UTB}, who build structures out of found tree branches.
Curved or arbitrarily shaped rods could be another direction for future
research.

\endgroup

\bibliographystyle{eg-alpha-doi}  
\bibliography{references}

\end{document}
